\documentclass[aps,amsmath,amssymb,reprint]{revtex4-2}
\usepackage{graphicx}
\usepackage[utf8]{inputenc}
\usepackage[T1]{fontenc}
\usepackage[pdfstartview=FitH, pdfpagemode=None, pdfpagelayout=OneColumn, colorlinks=true, linkcolor=blue, citecolor = blue, urlcolor = blue]{hyperref}

\begin{document}
\title{Maximization of accelerated charge in multibunch-driven plasma wakefield accelerators}
\author{K.V. Lotov}\email{K.V.Lotov@inp.nsk.su}
\affiliation{
Budker Institute of Nuclear Physics, Novosibirsk, 630090, Russia
}%

\date{\today}

\begin{abstract}
Driving the plasma wakefield with a train of multiple particle bunches has the potential to significantly increase the accelerated charge, provided that the witness bunches also form a train.
In this scenario, the driver bunches can replenish the wave energy after the passage of the witness bunches, yielding comparable charges in both trains.
However, implementing this method at high wakefield amplitudes is complicated by the reshaping of the driver bunches as they reach radial equilibrium with the wave.
The witness bunches cannot preserve their quality during this process; nevertheless, they must load the wave to prevent it from reaching the amplitude at which the wakefield period elongates and the wave destroys the driver.
The proposed solution is to form an equilibrium train of initially inefficient driver bunches, which becomes efficient upon the injection of the witness bunches.
This concept is demonstrated using the example of a self-modulating driver beam.
Several factors favoring a high witness charge are identified: injecting additional ``phaser'' bunches at the zero-accelerating-field points, suppressing the formation of an electron halo around the plasma, and operating at the highest possible wakefield amplitudes.
\end{abstract}

\maketitle

\section{Introduction}

Plasma wakefield acceleration \cite{CAS-2019,RMP81-1229,IEEE-PS24-252} is rapidly advancing toward higher-energy \cite{MRE9-014001,PRL133-255001,PPCF58-034017} and higher-quality \cite{PRL126-214801,Nat.640-907} beams, promising even greater achievements \cite{NJP23-031101,PoP27-070602,PRR7-023118,PPCF63-125027,BLPI50-S715}.
Usually, in this method, a short laser pulse or particle bunch (the driver) excites a Langmuir wave (the  wakefield) in the plasma, within which a short lepton bunch (the witness) is accelerated over a long distance, following the driver like a surfer follows a motorboat.
The wakefield can also be excited using a resonant train of short laser pulses or particle beams \cite{PRL43-267,PRL54-693,PRL104-255003}.
This makes sense if it is impractical to produce a single short driver \cite{JPB47-234003,IPAC10-4395,PPCF53-014003}.

In the case of multi-bunch excitation, the plasma wave can, in principle, periodically extract energy from the driver bunches and deliver it to multiple witness bunches arranged in a train.
This method has the potential to provide a significantly higher charge of the accelerated beam, but its study is still in the early stages \cite{PRST-AB7-111301,PRR7-L012055}.

In this paper, we examine how to maximize the accelerated charge assuming that both the driver and the witness are trains of short particle bunches.
We also assume that the wave is linear or weakly nonlinear, and specifically that its amplitude is well below the wavebreaking field $E_0 = m c \omega_p / e$, where $m$ is the electron mass, $c$ is the speed of light, $e$ is the elementary charge, and $\omega_p = \sqrt{4 \pi n_0 e^2/m}$ is the plasma frequency determined by the plasma electron density $n_0$.
Otherwise, the resonant excitation of the wave would be limited by the elongation of the wave period \cite{PoP20-083119}.
We consider only witness bunches matched to the wave such that they experience a constant average accelerating field and exhibit a low final energy spread (limited by the precision of bunch shaping).
This implies that these bunches are narrow relative to the characteristic transverse scale of the wave $k_p^{-1} = c/\omega_p$.
Otherwise, the radial variation of the accelerating field would lead to a large energy spread \cite{PAcc17-171}.

In Sec.~\ref{s2-theory}, we formulate theoretical limits on the total witness charge imposed by energy conservation under different assumptions about the driver and witness bunch shapes.
These limits neglect the peculiarities of realistic beams and plasma waves.
Therefore, after introducing the research methods and notation in Sec.~\ref{s3-methods}, we use numerical simulations to study the factors affecting the accelerated charge.
In Sec.~\ref{s4-stability}, we discuss the transverse stability as a necessary condition for accelerating trains of high-quality witnesses.
Then, in Sec.~\ref{s5-optimized}, we focus on stable drivers produced via beam self-modulation:
we first demonstrate a reference case with a high witness beam charge and then analyze the steps taken to obtain this high charge.
The final Sec.~\ref{s6-summary} summarizes the findings of this study.

\begin{figure}[tb]
\includegraphics{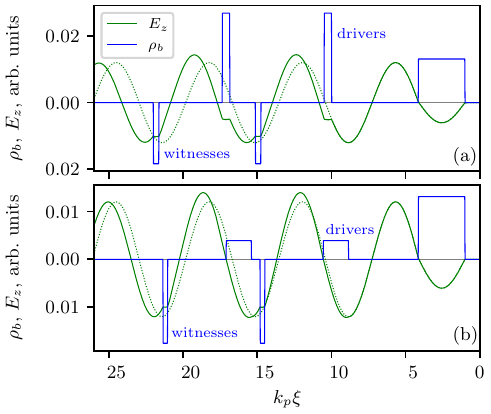}
\caption{ Illustration of idealized bunch positioning with short (a) and long (b) driver bunches: on-axis profiles of beam charge density $\rho_b$ and longitudinal electric field $E_z$.
For clarity, the driver and witness bunches are assigned opposite charges.
The dotted line represents the electric field generated solely by the first driver bunch.}
\label{fig01-theory}
\end{figure}

\section{Theoretical limits}
\label{s2-theory}

If we can create bunches of arbitrary shape and charge, then the only limitation on the witness charge is imposed by the energy conservation law.
In this idealized scheme, one part of the driver (with charge $Q_0$) generates the wave of the required amplitude, while the other part (consisting of several bunches with a total charge $Q_d$) replenishes the wave energy extracted by several witness bunches with a total charge $Q_w$ [Fig.\,\ref{fig01-theory}(a)].
Let $E_w$ denote the (common) accelerating field for all witness bunches, and $E_d$ the average decelerating field for the replenishing driver bunches.
In the absence of energy dissipation in the plasma, the ratio $R_q$ of the total witness charge to the total driver charge follows from the energy conservation law $Q_w E_w = Q_d E_d$:
\begin{equation}
    R_q = \frac{Q_w}{Q_d + Q_0} = \frac{E_d}{E_w (1 + Q_0/Q_d)}.
\end{equation}
If there are many driver bunches that replenish the energy, the fraction of the driver charge required to excite the wave can be relatively small, $Q_0 \ll Q_d$.
In this case, the witness charge can even exceed the driver charge if $E_d > E_w$.

When the generation of short, high-current driver bunches is challenging, an alternative scenario applies: the driver current is limited, and the most efficient wave excitation corresponds to driver bunches that completely fill the decelerating and focusing phases of the wave \cite{AIP396-75} [Fig.\,\ref{fig01-theory}(b)].
If the bunch occupies a quarter of the wave period and does not significantly perturb the field amplitude $E_\text{max}$, then for a harmonic wave $E_d \approx 2 E_\text{max} / \pi$, yielding the charge ratio
\begin{equation}\label{e2}
    R_q \approx \frac{2}{\pi}\frac{E_\text{max}}{E_w (1 + Q_0/Q_d)}.
\end{equation}

Several methods exist for generating a bunched beam while preserving most of its initial charge \cite{RMP94-025006,JINST17-P05036,RuPAC16-303}.
To date, however, none of them has yet been experimentally demonstrated for a high-charge, high-energy (GeV-scale) driver beam.
A more reliable approach is to slice individual bunches from a long beam with a smoothly varying current \cite{PFB5-2714,PPR31-292,AIP1299-467,PRST-AB13-052803,NIMA-916-195,PRAB23-072803, PRL104-255003,PoP22-103110}.
In this case at least 3/4 of the beam charge must be discarded.
Therefore, the witness charge should be compared to the initial driver charge, which is now $4(Q_d + Q_0)$.
The limit on the witness charge becomes 4 times lower:
\begin{equation}\label{e3}
    R_q \approx \frac{1}{2 \pi}\frac{E_\text{max}}{E_w (1 + Q_0/Q_d)}.
\end{equation}

The acceleration rate also plays a critical role in plasma wakefield accelerators, so the witness bunches are typically positioned in a strong field near its maxima.
When $E_w \approx E_\text{max}$, the limiting value of the charge ratio is
\begin{equation}\label{e4}
    R_q \approx \frac{1}{2 \pi (1 + Q_0/Q_d)} < \frac{1}{2\pi},
\end{equation}
subject to all the assumptions stated above.

\section{Methods and notation}
\label{s3-methods}

We validate the main findings of this work using numerical simulations performed with the quasistatic 2d3v (axisymmetric) code LCODE \cite{PRST-AB6-061301,NIMA-829-350}.
The code calculates the plasma response in the co-moving window $(\xi, r)$, where $\xi = z - ct$, and $(r, \varphi, z)$ are standard cylindrical coordinates with the $\vec z$-axis aligned with the direction of beam propagation.
Since the beams and processes under study are axisymmetric, full three-dimensional simulations are  unnecessary.

Except for the illustrative cases in Figs.~\ref{fig01-theory}--\ref{fig03-preformed3}, the simulation grid steps are $\Delta \xi = \Delta r = 0.01 k_p^{-1}$, and the step for updating the plasma state is $\Delta z = 200 k_p^{-1}$.
Each plasma component (electrons and ions) is represented by 10 radius-weighted macroparticles per radial cell $\Delta r$.
Both driver and witness macroparticles carry a charge of 2.5\,fC (corresponding to $\pm 1.56 \times 10^4 e$).
Unless otherwise specified, the simulation domain has a radius of $8 k_p^{-1}$ and a length equal to the beam length.

It is convenient to characterize the combined action of the electric ($\vec E$) and magnetic ($\vec B$) fields on the beams by the dimensionless wakefield potential $\tilde \Phi$, whose gradient determines the force on an axially moving relativistic unit charge:
\begin{equation}
    \tilde \Phi = - \frac{e}{mc^2} \int E_z d \xi = - \frac{e}{mc^2} \int (E_r - B_\varphi) d r.
\end{equation}
The radial dependence of $\tilde \Phi$ is usually monotonic, although this is not always the case for complex wakefield patterns.
Therefore, the on-axis wakefield potential $\Phi \equiv \tilde \Phi|_{r=0}$ indicates (but does not precisely determine) whether the wakefield is focusing or defocusing.
We also use $\Phi_m$ to denote the amplitude of potential oscillations, and $\Phi_\text{max}$ to denote the maximum amplitude reached at a given $z$.

The contribution of individual beam slices to wakefield formation can be characterized and visualized by the dimensionless effective current \cite{PoP32-033102, PPCF64-075003}:
\begin{equation}\label{e6}
    I_\text{eff} (\xi, z) = \frac{k_p^2}{n_0} \int_0^\infty K_0 (k_p r) n_b (r, \xi, z)\, r \, dr,
\end{equation}
where $n_b$ is the beam density and $K_0$ is the modified Bessel function of the second kind.
The effective current determines the on-axis wakefield potential in a linearly responding plasma:
\begin{equation}
    \Phi (\xi, z) = - k_p \int_\xi^\infty \, I_\text{eff} (\xi', z) \, \sin \bigl( k_p (\xi' - \xi)\bigr) \, d\xi',
\end{equation}
thereby allowing the beam bunches to be represented on line plots, where their radial structure is averaged and only their efficiency is retained.
We also introduce the effective current normalized to the peak beam density $n_{b0}$:
\begin{equation}
    \tilde I_\text{eff} = \frac{n_{b0}}{n_0} I_\text{eff},
\end{equation}
and the normalized effective current $\bar I_\text{eff}$, further averaged over the wakefield period.

The integral of the effective current, the effective beam charge, quantifies the beam's compression and total charge loss:
\begin{equation}
    Q_\text{eff} (z) = \int I_\text{eff} (\xi, z) \, d\xi.
\end{equation}

The on-axis electric field, averaged over the witness acceleration path (from $z_w$ to $z_m$), determines the optimal positioning of the witness bunches:
\begin{equation}
    \langle E_z \rangle = \frac{1}{z_m - z_w} \int_{z_w}^{z_m} E_z (\xi, 0, z) \, dz.
\end{equation}
This function of $\xi$ is oscillatory, and we denote its amplitude --- measured at the local minima --- by $E_m$.
The minima are of particular interest because witness electrons are accelerated in regions where $\langle E_z \rangle < 0$.

The wave energy content is quantified by the energy flux along the simulation window:
\begin{multline}
    \Psi (\xi, z) = \int \left( \frac{E^2 + B^2}{8\pi} - \frac{1}{4\pi} \left[ \vec{E} \times \vec{B} \right]_z \right) c \, dS \\
    + \sum_j (\gamma_j - 1) m_j c^2,
\end{multline}
where $m_j$ and $\gamma_j$ are the mass and relativistic factor of the plasma particles, respectively; the integration is performed over the transverse cross-section at $\xi$; the summation runs over all plasma particles intersecting this cross section in unit time.
The natural unit for this energy flux is
\begin{equation}
    \Psi_0 = \frac{m^2 c^5}{4 \pi e^2}.
\end{equation}
The quantity $\Psi/c$ also represents the linear energy density deposited by the beam in the plasma.
A flux of $\Psi_0/c$ corresponds to a linear energy density of 2\,J/m.

\section{Beam stability}
\label{s4-stability}

The problem of efficient energy transfer is directly related to the issue of radial beam equilibrium.
The beams in plasma experience a transverse force that is generally of the same order of magnitude as the longitudinal (accelerating or decelerating) force.
Due to the large relativistic factor $\gamma_b$, the beams respond to the transverse force much faster than to the longitudinal force.
Depending on the sign of the transverse force, parts of the beam are either expelled from the wave or come into radial equilibrium with it \cite{PoP5-785,NIMA-410-461}.
The transverse relaxation time is shorter than the energy exchange timescale by a factor of about $\sqrt{\gamma_b}$ \cite{PPR31-292}.
Consequently, the establishment of transverse equilibrium always precedes the process of energy exchange.

However, the bunches change shape as they approach transverse equilibrium.
Contributions of individual beam slices to the excitation or damping of the wave also change.
As a result, the amplitude and phase of the excited wakefield change.
The problem is compounded by the fact that the equilibrium radial profiles of the beam density are rather exotic \cite{PoP24-023119,features} and are difficult to obtain by any means other than self-consistent evolution within the plasma itself.
Therefore, any sequence of driver and witness bunches prepared externally is not in radial equilibrium with the excited wave and is thus subject to modification in the plasma.

\begin{figure}[tb]
\includegraphics{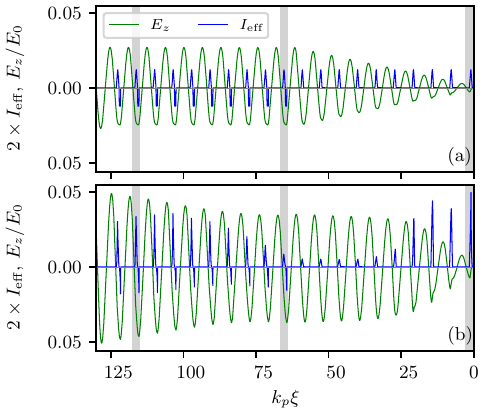}
\caption{ Illustration of how the beams can change as they approach the radial equilibrium: effective current $I_\text{eff}$ and on-axis electric field $E_z$ (a) near the entrance to the plasma and (b) after reaching the radial equilibrium at $z_1 = 5000 k_p^{-1}$.
Shaded areas indicate the bunches shown in Fig.\,\ref{fig03-preformed3}.}
\label{fig02-preformed1}
\end{figure}

\begin{figure}[tb]
\includegraphics{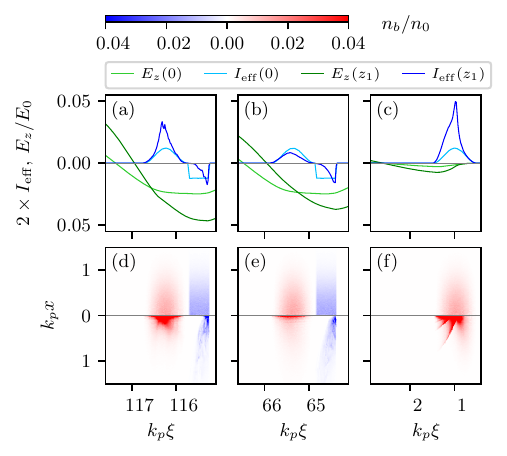}
\caption{ Evolution of wakefields and bunch shapes across the bunch train as the beams approach radial equilibrium. (Top row) effective current $I_\text{eff}$ and on-axis electric field $E_z$; (bottom row) beam density $n_b$ at the plasma entrance (upper panels) and at equilibrium ($z_1 = 5000 k_p^{-1}$, lower panels).}
\label{fig03-preformed3}
\end{figure}

Figures~\ref{fig02-preformed1} and~\ref{fig03-preformed3} illustrate this problem.
Here, we simulate a manually prepared bunch train in which the first half (10 bunches) of the driver excites the wave, while the second half replenishes the wave energy extracted by the train of short witness bunches.
For clarity, the driver and the witness are assigned opposite charges, although this does not matter because the linear plasma response to the beams is charge-symmetric.
Initially [Fig.\,\ref{fig02-preformed1}(a)], all driver bunches are identical and reside entirely within the focusing and decelerating phase of the wave, as close as possible to the maximum decelerating field.
Since the period of the driven wave slightly changes as the wave amplitude increases \cite{PPCF60-024002,AIP396-75}, the bunch-to-bunch spacing varies within the first half of the driver beam.
The bunches in the second half of the driver train are equidistant.
The witness bunches are also identical, equidistant, and reside in the same accelerating field.
Their charge is chosen such that the wakefield amplitude remains constant.
Initially, all bunches have the same radius $\sigma_r = k_p^{-1}$, relativistic factor $\gamma_b = 1000$, and angular spread $\delta \alpha = 10^{-5}$.

While reaching radial equilibrium, all bunches evolve [Fig.\,\ref{fig02-preformed1}(b)].
Some bunches are compressed by the wakefield, increasing their effective current [Fig.\,\ref{fig03-preformed3}].
Others, including the witness bunches, are partially defocused, losing a significant portion of their charge.
The wakefield phase and amplitude change, and the perfect matching between the bunches and the wave disappears.

To summarize this part, it is practically impossible to form a perfect sequence of driver and witness bunches before they enter the plasma in such a way that it remains perfect inside the plasma.
Hypothetically, it is not proven impossible to create a bunch train that transforms into the desired one as it reaches equilibrium.
However, this is extremely challenging even in numerical simulations.

Fortunately, there is a way to form a good equilibrium bunch train inside the plasma.
This is the so-called seeded self-modulation (SSM), a process in which perturbations of the beam radius couple to the excited wakefield and amplify each other \cite{EPAC98-806,PRL104-255003,PoP22-103110}.
The SSM transforms a long beam into a resonant train of short bunches \cite{PRL122-054802,PRL122-054801,PRL125-264801,PRL126-164802,PRAB24-101301}.
Each bunch is focused by the wakefield and is in radial equilibrium with the wave \cite{features}.
The bunches are not strictly equal and equidistant, as required to maintain synchronism with the growing wave \cite{PRL122-054801,PRL125-264801,PRAB24-101301}.
The positions of individual bunches --- and, consequently, the wakefield phase --- are dictated by a seed perturbation, which can be generated by a relativistic ionization front \cite{PRL126-164802}, the leading edge of the beam \cite{PPCF56-084014,PRL112-045001,PRL120-144802,PPCF61-045012},
a short electron beam \cite{PPCF62-115025,PRL129-024802},
or a laser pulse \cite{PoP20-103111}.
The remaining beam charge can be controlled by adjusting the longitudinal plasma density distribution during the self-modulation stage \cite{PoP22-103110}.
Compared to a uniform plasma, a plasma with a step-like \cite{PoP18-024501,PoP18-103101,PPCF62-115025,Symmetry14-1680}
or gradual (gradient) \cite{NIMA-829-63,Nat.561-363,PRL125-264801,PPCF62-125023,PRAB24-101301}
density increase along the propagation direction allows the bunches to preserve more charge and drive a stronger wakefield.
However, the bunches formed by the SSM are not perfect in that portions of them reside in the accelerating phase of the wave \cite{PoP22-103110,features}.
Varying the plasma density profiles allows one to control the bunch shape --- including the charge fraction in the accelerating phase --- and to form bunch trains that drive the wave with different efficiencies or generate waves with different phase velocities \cite{PRL107-145003,PoP19-010703, JPCS1067-042009,PPCF62-115025,PPCF63-125027,PoP32-033102}.
A plasma density profile capable of producing an ideal bunch train, one containing no particles in the accelerating phase, has not yet been devised.
If such a profile exists, the limit on its efficiency is given by Eq.~\eqref{e3} or Eq.~\eqref{e4}.

Next, we focus on driver bunch trains produced with the SSM.

\begin{table}[tb]
\begin{center}
\caption{Parameters of the reference case}\label{t1}
\begin{tabular}{lll}\hline\hline
  Parameter, notation && Value \\ \hline
  \textit{Plasma:} && \\
  Initial density, $n_0$ && $7 \times 10^{14}\,\text{cm}^{-3}$ \\
  Density rise, $\delta n_s$ && $0.053 n_0$ \\
  Total length, $z_m$ && 20\,m \\
  Length scale of density rise, $z_s$ && 4\,m \\
  Plasma ion mass number && 85 \\
  \textit{Proton driver:} && \\
  Population (charge, $Q_b$) && $3\times 10^{11}$ ($Q_b = 48$\,nC) \\
  Length, $\sigma_{zb}$ && 5.1\,cm \\
  Radius, $\sigma_{rb}$ && 0.2\,mm \\
  Peak density, $n_{b0}$ && $0.0133 n_0$ \\
  Energy && 400\,GeV \\
  Energy spread && 0.035\,\% \\
  Normalized emittance && 2.5\,mm\,mrad \\
  Relative location of leading edge, $\xi_c$ && $2 \sigma_{zb}$ \\
  \textit{Witness bunches:} && \\
  Total charge, $Q_w$ && 1.11\,nC \\
  Radius, $\sigma_{re}$ && $5.4\,\mu$m \\
  Initial energy, $W_w$ && 190\,MeV \\
  Energy spread && 0\,\% \\
  Normalized emittance && 2\,mm\,mrad \\
  Injection point, $z_w$ && 14\,m \\
  Average accelerating field, $E_w$ && $0.3 E_0 \approx 765\,\text{MV}/\text{m}$ \\
  Threshold field amplitude, $E_a$ && $0.36 E_0$ \\ \hline
  \hline
 \end{tabular}\end{center}
\end{table}

\section{Optimized acceleration}
\label{s5-optimized}

\subsection{Reference case}
\label{s-refcase}

First, we describe a reference case featuring a high accelerated charge distributed among multiple witness bunches, and then, in the following subsections, we detail the steps required to obtain it.
The initial parameters of the driver beam in the reference case correspond to those of the proton beam in recent AWAKE experiments (Table~\ref{t1}) \cite{PRL134-155001, 2602.16624, PRAB29-083601}.
These experiments \cite{NIMA-829-3,NIMA-829-76,Symmetry14-1680}, conducted at CERN, currently provide the largest amount of experimental data on the interaction of a high-energy bunched beam with plasma, so it is natural to use them as a reference point.
Looking ahead, we should note that this set of parameters is not the most advantageous for demonstrating the capabilities of multi-bunch witnesses, as it requires a large charge $Q_0$ to excite the wave.
However, we deliberately choose it to yield results experimentally achievable in the foreseeable future.

\begin{figure}[tb]
\includegraphics{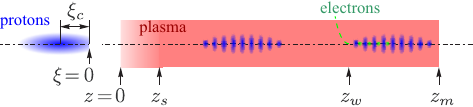}
\caption{ Schematic of the reference case.}
\label{fig04-setup}
\end{figure}

The driver bunch train is produced by the self-modulation seeded by a relativistic ionization front.
In simulations, we emulate this front by sharply cutting off the head of the beam, since its effect on the plasma wave is the same.
The initial beam shape is
\begin{equation}
    n_b (r, \xi) = \frac{n_{b0}}{2} \, e^{-r^2/(2 \sigma_{rb}^2)} \left[ 1 + \cos \left( \frac{\sqrt{\pi}(\xi+\xi_c)}{\sqrt{2} \sigma_{zb}}  \right) \right]
\end{equation}
for
\begin{equation}
    - \sigma_{zb} \sqrt{2\pi} - \xi_c  < \xi < 0
\end{equation}
and is zero otherwise (Fig.\,\ref{fig04-setup}).

\begin{figure}[tb]
\includegraphics{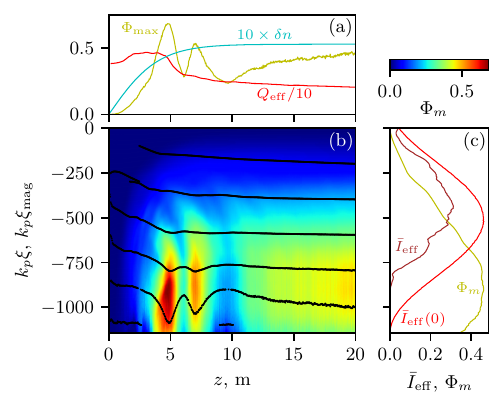}
\caption{ Features of the unloaded reference case.
(a) The plasma density profile, represented by the density increment $\delta n = (n(z)-n_0)/n_0$, the maximum wakefield potential $\Phi_\text{max}$, and the effective beam charge $Q_\text{eff}$ as functions of the propagation distance $z$.
(b) Map of the wakefield potential amplitude $\Phi_m(\xi,z)$ with several lines $\xi_\text{mag} (z)$ superimposed on it to illustrate the change in the wave phase.
(c) Variation along the beam of the average effective current $\bar I_\text{eff}$ and the wakefield potential amplitude $\Phi_m$ at $z=20$\,m, as well as the effective current at $z=0$ [labeled as $\bar I_\text{eff} (0)$].}
\label{fig05-reference}
\end{figure}

The driver beam propagates through a 20-meter-long plasma section, whose density initially varies and then gradually approaches a constant value.
The density profile is optimized to maximize the number of beam particles that remain in the equilibrium beam behind the point where the wakefield amplitude exceeds a certain threshold.
Specifically, we maximize the quantity
\begin{equation}\label{e14}
    A = \int_{-\infty}^{\xi_{0.3}} I_\text{eff} (\xi, z_m) \, d \xi,
\end{equation}
where $\xi_{0.3}$ is the maximum value of $\xi$ for which the inequality $\Phi(\xi) > E_w/E_0 = 0.3$ holds.
By choosing this objective function, we maximize the driver charge $Q_d$ available to replenish the wakefield at large amplitudes.
When optimizing the density profile, we relaxed the constraints of the AWAKE experiment, in which the implementable density profiles are piecewise-constant \cite{SPSC-SR-337} or linear \cite{JPD51-025203}.
The highest value of the objective function is achieved for the plasma density profile
\begin{equation}
    n (z) = n_0 + \delta n_s \tanh (z / z_s)
\end{equation}
where $\delta n_s = 0.053 n_0$ and $z_s = 4$\,m [Fig.\,\ref{fig05-reference}(a)].
Piecewise-linear density profiles, similar to those considered in Ref.~\cite{PPCF62-115025}, yielded lower values of the objective function.
For consistency of the length scale along the plasma column, we retain the notation $k_p$ for the wave number corresponding to the initial plasma density $n_0$.

Figure~\ref{fig05-reference} shows how the reference proton beam evolves in the absence of electron bunches.
The effective beam charge $Q_\text{eff}$ first increases slightly because of beam compression by the focusing wakefield, then decreases as the beam loses defocused particles, and finally, after $z \sim 10$\,m, slowly approaches some reduced value [Fig.\,\ref{fig05-reference}(a)].
The overall reduction in the effective charge is less than that of the actual charge, since the radial distribution of the beam density becomes more peaked \cite{features}, allowing individual beam slices to excite the wakefield more efficiently than in their initial Gaussian state.

With this particular profile of the plasma density, a larger fraction of the beam charge is lost in the tail of the beam, as evidenced by the effective current plots in Fig.\,\ref{fig05-reference}(c).
This indicates that the bunches are smaller in this region.
However, the lower charge of these bunches is compensated by their higher efficiency, enabling the wakefield amplitude $\Phi_m$ to continue growing almost linearly along the beam.

To characterize the wakefield phase behavior, we track the coordinates $\xi_\text{loc} (z)$ of the zero-field points on the axis where $E_z = 0$ and $\partial E_z / \partial \xi > 0$.
To relate the phase variations to the changes in wakefield amplitude, in Fig.\,\ref{fig05-reference}(b) we plot the quantity
\begin{equation}\label{e16}
    \xi_\text{mag} (z) = \frac{100}{\pi} \bigl( \xi_\text{loc} (z) - \xi_\text{loc} (z_m) \bigr) + \xi_\text{loc} (z_m)
\end{equation}
instead of $\xi_\text{loc} (z)$, thereby magnifying small phase changes and making them easier to see.
The coefficients in Eq.~\eqref{e16} are chosen so that the distance between the black lines in Fig.\,\ref{fig05-reference}(b) corresponds to the phase change of $2\pi$, with the rightmost point of each line indicating the exact position along $\xi$ at which the phase is monitored.
After the beam has traveled about 15\,m, the wakefield phase stabilizes along the entire length of the beam, except for its tail.
Beyond this point, the self-modulation can be considered complete, and the wakefield becomes suitable for accelerating precisely positioned witness bunches.
However, the wakefield amplitude continues to vary [Fig.\,\ref{fig05-reference}(a)], indicating that full stabilization of the wakefield is not possible and should not be awaited before placing the witnesses.
We choose $z_w = 14$\,m as the witness injection point.

The initial energy $W_w$ of the witness electrons is chosen such that their velocity is close to the phase velocity of the wakefield near the injection point.
The initial witness radius is matched to the focusing force of plasma ions \cite{PRAB21-011301}; that is, the witness radius does not oscillate if the witness locally creates its own bubble (a region free of plasma electrons).
The shape and current of all witness bunches are matched to the wave to minimize the final energy spread.
For both linear and highly nonlinear waves, the energy spread is minimal if the longitudinal profile of the witness current is close to a trapezoid \cite{PAcc22-81,PoP12-053105,PoP16-056705,PoP30-053108}.
If the witness is short, the trapezoid reduces to a rectangle of constant current.

\begin{figure}[tb]
\includegraphics{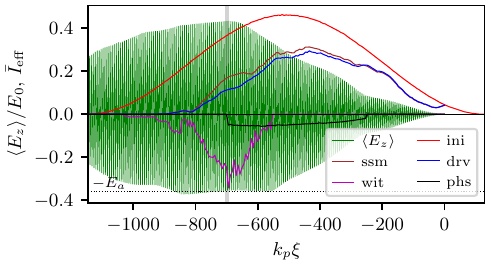}
\caption{ Loaded reference case: the average electric field $\langle E_z \rangle$ and the averaged effective currents $\bar I_\text{eff}$ of the proton beam at $z=0$ (ini), at $z=z_w$ (ssm), and at $z=z_m$ (drv), of the witness bunches at $z=z_m$ (wit), and of the phasers at $z=z_m$ (phs).
The dotted line indicates the threshold field $(-E_a)$, and the vertical grey bar marks the area detailed in Fig.\,\ref{fig07-best2}.}
\label{fig06-best1}
\end{figure}

\begin{figure}[tb]
\includegraphics{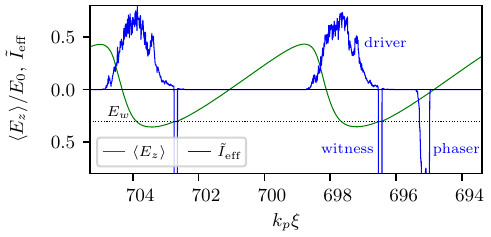}
\caption{ Normalized effective current $\tilde I_\text{eff}$ of individual bunches at $z=z_m$ and the average electric field $\langle E_z \rangle$ in the area marked in Fig.\,\ref{fig06-best1}. }
\label{fig07-best2}
\end{figure}

We construct the train of witness bunches by tracing the average electric field $\langle E_z \rangle$ backward from the beam front using the following algorithm:
\begin{enumerate}
  \item We wait until the wakefield amplitude $E_m$ exceeds the threshold value $E_a$ (Fig.\,\ref{fig06-best1}).
      This occurs at $k_p \xi \approx -560$.
  \item We begin injecting a constant-current witness bunch during the next wakefield period when $\langle E_z \rangle$ reaches the designed accelerating field $(-E_w)$ (Fig.\,\ref{fig07-best2}).
      The bunch current is chosen to ensure a constant average field $\langle E_z \rangle$ throughout the bunch.
      This current is proportional to $\partial_\xi \langle E_z \rangle$ at the bunch front.
      Once determined, the proportionality coefficient remains fixed for all witness bunches.
  \item We choose the witness length such that the wave amplitude $E_m$ is restored to $E_a$ after the witness has taken energy from the wave and the proton bunch has replenished it.
      This is achieved by performing an additional iteration in which the next wakefield period is calculated neglecting the next witness bunch.
      This is possible because quasistatic codes can resume the run starting from any $\xi$ \cite{PPNL21-316}.
  \item We repeat Steps 2-3 until the wave amplitude $E_m$ fails to reach $E_a$ after the witness bunch passage.
  \item When the above occurs (at $k_p \xi \approx -850$), we continue to place witnesses as described in Step 2, but set the witness length proportional to $(E_m-E_w)$.
      This final step is required to utilize the wakefield energy as efficiently as possible.
\end{enumerate}

To increase the total witness charge, we inject additional short electron bunches into the wakefield in the phases of zero accelerating field and strong focusing.
We call them phasers and describe their role in Sec.\,\ref{s-phasers}.

\begin{figure}[tb]
\includegraphics{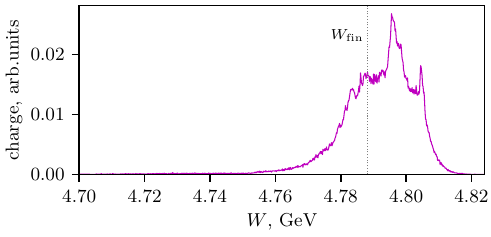}
\caption{ Final energy spectrum of the witness beam.
The dashed line shows the expected energy $W_\text{fin}$. }
\label{fig08-best3}
\end{figure}

In the reference case, the total charge of the witness bunches is 1.11\,nC.
The electrons are accelerated to the target energy
\begin{equation}
    W_\text{fin} = W_w + e E_w (z_m - z_w) \approx 4.788\,\text{GeV}
\end{equation}
with a small final energy spread of 0.4\% (Fig.\,\ref{fig08-best3}).
The witness bunches are nearly equidistant, with the bunch-to-bunch distance varying within the interval $(6.22 \pm 0.02) k_p^{-1}$.
The charge of individual bunches fluctuates because the bunch length takes on discrete values that are multiples of the grid step $\Delta \xi$ [``wit'' line in Fig.\,\ref{fig06-best1}].
The largest witness bunch has a charge of 38\,pC.
More than half of the initial driver charge is used to bring the wave to a high amplitude: $Q_0 \approx 0.6 Q_b$.
Therefore, the theoretical limit \eqref{e4} for the witness charge is about 3.2\,nC.
The actual witness charge is below the theoretical limit, primarily because the drive beam is bunched inefficiently in the tail region, where less than a quarter of the initial charge remains in the bunches.

Next, we analyze the factors that led to the achieved witness charge.

\begin{figure}[tb]
\includegraphics{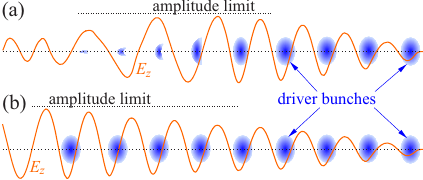}
\caption{ Illustration of the amplitude limit problem: rapid (a) and slow (b) wave growth. }
\label{fig09-limit}
\end{figure}

\subsection{Objective function for self-modulation}

It is easier to work with multiple bunches in the driver or witness if the wave is linear and the accelerating field in it is much less than $E_0$.
However, it is often necessary to maximize the accelerating gradient and operate in a weakly nonlinear regime, when the wakefield amplitude approaches the limit imposed by the nonlinear elongation of the wave period \cite{PoP20-083119}.
This leads to a problem illustrated in Fig.\,\ref{fig09-limit}.
If the front part of the bunch train rapidly drives the wave to the limiting amplitude, the rear part finds itself in a wave with an elongated, amplitude-dependent period \cite{PoP20-083119,PPCF63-125027}, which is prone to quick wavebreaking \cite{NIMA-829-3}.
The bunches get destroyed there and become unsuitable for wave replenishment [Fig.\,\ref{fig09-limit}(a)].
If the beam excites the wave slowly [Fig.\,\ref{fig09-limit}(b)], a high-amplitude wave remains behind the driver, but there are no driver particles at that location to replenish the wave when the witnesses take away its energy.
Therefore, a bunched beam capable of producing and regenerating a high-amplitude wave cannot survive in the plasma without witnesses that load the wave.

The witnesses can prevent the wave from reaching excessive amplitudes, but for this, they must appear in the plasma before the beam reaches transverse equilibrium.
For a pre-bunched beam, this is challenging because of the wave phase shifts during driver equilibration, as shown in Sec.\,\ref{s4-stability}.
If the beam becomes bunched due to self-modulation, the witness bunches cannot be injected into the wakefield until the self-modulation is complete, for another reason.
In this case, the witnesses generally shift to the bottoms of the potential wells, where they experience strong focusing and almost no acceleration, except in rare cases where a favorable interplay of nonlinear effects allows a witness to fall into a strong accelerating field \cite{PoP21-123116,NIMA-829-63,PRAB24-011301,PRAB22-031301}.

However, it is possible to circumvent the above problems by creating a bunch train in which the bunches are initially inefficient and contain both decelerated and accelerated parts.
Such a bunch train can reach transverse equilibrium in the plasma without driving the wave to an excessive amplitude.
When the wave is loaded with witness bunches, its phase shifts, causing the bunches to become more efficient at replenishing the wave energy.
It is this approach that enabled the wave recovery in Ref.\,\cite{PRR7-L012055}.

The phase dynamics can be visualized using the technique of complex wave amplitudes for linearly responding plasmas \cite{PoP22-103110,PPCF60-024002,PPCF64-075003}.
We characterize the wave by the complex wakefield potential $\Phi_c$ and its complex amplitude $\Phi_A$, which contain information about both the amplitude and phase of the wakefield:
\begin{equation}
    \Phi_c(\xi) = \Phi_A (\xi) e^{i k_p' \xi},
\end{equation}
\begin{equation}
    \Phi_A (\xi) = -i k_p' \int_\xi^\infty I_\text{eff} (\xi') e^{-i k_p' \xi'} \, d\xi',
\end{equation}
\begin{equation}
    \Phi (\xi) = \text{Re} \bigl( \Phi_c \bigr), \quad
    E_z(r=0,\xi) = k_p' \text{Im} \bigl( \Phi_c \bigr),
\end{equation}
where the wavenumber $k_p'$ corresponds to the local plasma density.
In beam-free regions, $\Phi_A (\xi) = \text{const}$.

\begin{figure}[tb]
\includegraphics{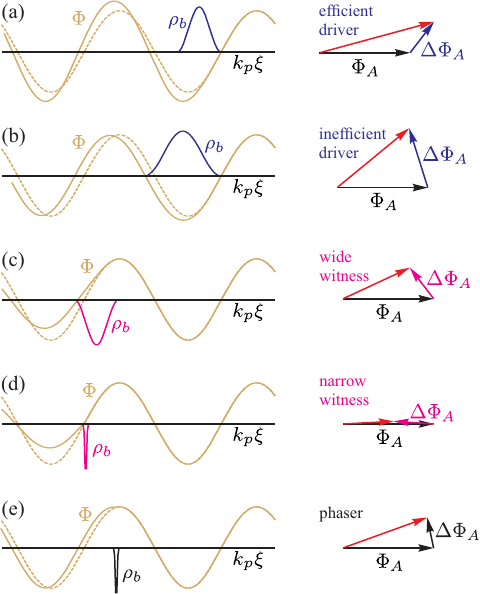}
\caption{ Changes in the complex amplitude $\Phi_A$ (right) and wakefield potential $\Phi$ (left) induced by individual bunches.
Solid and dashed lines show the wave with and without the bunches, respectively.}
\label{fig10-arrows}
\end{figure}

Each bunch contributes to the complex amplitude $\Phi_A$, as illustrated by a vector sum in the complex plane (Fig.\,\ref{fig10-arrows}).
Efficient driver bunches are located in the focusing and decelerating phases of the wave.
Their contribution increases the wave amplitude and shifts the wave phase backward [Fig.\,\ref{fig10-arrows}(a)]; that is, the argument of $\Phi_A$ increases, and the field maxima shift backward along $\xi$.
An inefficient driver bunch that completely fills the focusing interval does not increase the wave amplitude but efficiently shifts the wave phase backward [Fig.\,\ref{fig10-arrows}(b)].
A wide witness bunch reduces the wave amplitude and shifts the phase backward [Fig.\,\ref{fig10-arrows}(c)].
A narrow witness bunch located in the strongest accelerating field reduces the wave amplitude but has almost no effect on its phase [Fig.\,\ref{fig10-arrows}(d)].

\begin{figure}[tb]
\includegraphics{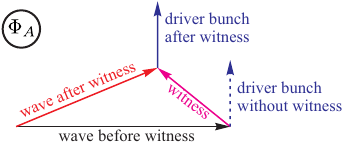}
\caption{ Change in the driver bunch contribution to the complex wakefield amplitude $\Phi_A$ induced by a witness bunch. }
\label{fig11-triangle}
\end{figure}

If a witness bunch loads the wave, this changes the wave phase such that the nearby inefficient driver bunches become more efficient and begin to replenish the wave (Fig.\,\ref{fig11-triangle}).
The bunch contribution to the wave changes from the pattern in Fig.\,\ref{fig10-arrows}(b) to the pattern in Fig.\,\ref{fig10-arrows}(a).

\begin{figure}[tb]
\includegraphics{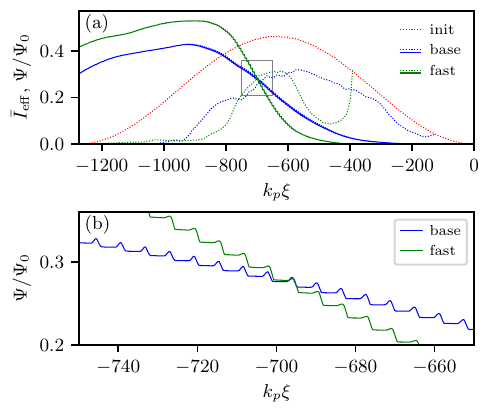}
\caption{ Energy content of the wave $\Psi$ (solid lines) and average effective current $\bar I_\text{eff}$ (dotted lines) before the witness injection point $z = z_w$ for the reference case (base) and the case with fast wakefield growth (fast).
The initial effective current (init) is also shown.
Fragment (b) provides a magnified view of the wave energies in the area marked by the gray rectangle in fragment (a).}
\label{fig12-aim}
\end{figure}

The efficiency of individual bunches formed as a result of self-modulation is controlled by adjusting the longitudinal profile of the plasma density.
When searching for the reference case, we directed the optimization algorithm toward large, inefficient bunches by choosing the objective function in the form \eqref{e14}.
If the density profile is optimized differently, the accelerated charge is lower, even though the wave energy content is higher.
This is illustrated by Fig.\,\ref{fig12-aim}(a), which compares the reference case to a variant optimized for maximum wakefield amplitude at $z=z_w$.
In this variant, the leading edge of the proton beam is closer to its midplane ($\xi_c = \sigma_{zb}$) to provide a stronger seed for self-modulation, and the plasma density jumps sharply from $n_0$ to $1.035 n_0$ at $z=1.1$\,m.
The energy content of the wave, determined by the energy flux $\Psi(\xi, z_w)$, is 23\% higher.
However, the witness charge, distributed according to the algorithm described in Sec.~\ref{s-refcase}, is 30\% lower.
A close-up view of the energy flux growth [Fig.\,\ref{fig12-aim}(b)] confirms the origin of this difference.
In both variants, the energy flux increases in steps corresponding to individual proton bunches.
Each step consists of an ascent followed by a minor descent.
In the reference case, the descent is much greater, indicating the presence of a large accelerated charge in the bunches.

\subsection{Phasers}
\label{s-phasers}

As follows from Fig.\,\ref{fig10-arrows}(d), if the witness bunch is located near the maximum of the accelerating field, it does not shift the wakefield phase, and consequently, does not initiate the wave recovery process.
This issue can be solved with the help of phasers --- additional electron bunches injected into the points of zero accelerating field and strongest focusing (Fig.\,\ref{fig07-best2}).
The phasers do not absorb the wave energy but efficiently shift the wave phase backward [Fig.\,\ref{fig10-arrows}(e)], thereby increasing the efficiency of the subsequent driver bunches.

The phaser parameters affect the maximum possible accelerated charge and require optimization.
In the reference case, the phasers of length $\Delta \xi$ and charge proportional to the wave amplitude are placed in each wave period within the interval $k_p \xi \in [-696, -247]$ (Fig.\,\ref{fig06-best1}).
The phasers are injected at the points where $E_z(\xi, 0, z_w) = 0$.
Their longitudinal shape is irrelevant, since their initial length equals one grid step.
Other phaser parameters are the same as for the witness bunches.
The total charge of the phasers is about half the total witness charge.

\begin{figure}[tb]
\includegraphics{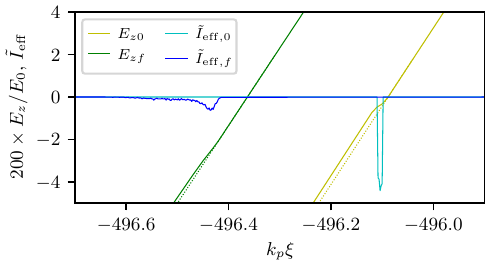}
\caption{ Evolution of a phaser: its normalized effective current $\tilde I_\text{eff}$ and the electric field $E_z$ in its vicinity at $z=z_w$ (phaser injection point, index ``0'') and at $z=z_m$ (end of interaction, index ``f'').
Dotted lines show the electric field in the absence of the phaser, and the difference between the dotted and solid lines indicates the induced phase shift.  }
\label{fig13-phaser}
\end{figure}

\begin{figure}[tb]
\includegraphics{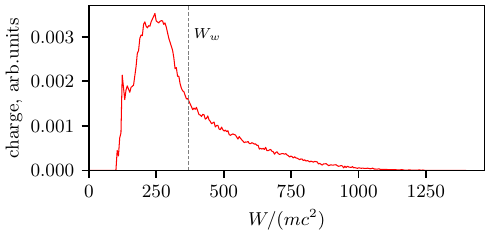}
\caption{ Final energy spectrum of the phasers.
The dashed line shows the initial phaser energy $W_w$. }
\label{fig14-pspectrum}
\end{figure}

Although the phasers are initially located at the points of zero accelerating field, they cannot stay there permanently because the phase velocity of the wave differs from the longitudinal velocity of phaser particles.
If the zero-field points shift backward, as in the reference case, the phasers follow them (Fig.\,\ref{fig13-phaser}), oscillating longitudinally near the bottoms of their potential wells.
The mechanism behind these oscillations is the same as in the case of witness bunch injection during driver self-modulation \cite{PoP21-123116,NIMA-829-63}.
If the local phase velocity of the wave is greater than the phaser velocity, the phaser shifts into the accelerating field, and its energy and velocity increase.
If the phaser moves faster than the wave, it enters the decelerating field and slows down.
The amplitude of these oscillations is minimized by choosing the phaser energy such that the phaser velocity matches the wakefield phase velocity near the phaser injection point.
In this case, even if the phase velocity of the wave eventually approaches or exceeds $c$, the phasers remain outside the regions of strong accelerating field and fail to acquire a high energy, comparable to that of the witness, over the interaction length (Fig.\,\ref{fig14-pspectrum}).
When moving backward in $\xi$, the phasers elongate, yet they continue to effectively shift the wave phase backward (Fig.\,\ref{fig13-phaser}).

\begin{figure}[tb]
\includegraphics{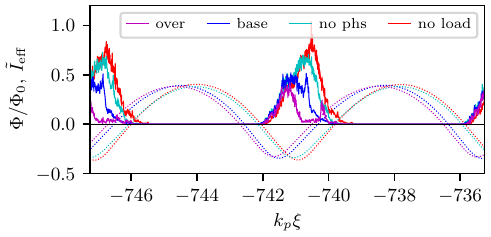}
\caption{ Destruction of driver bunches caused by wave phase shifts.
Normalized effective current $\tilde I_\text{eff}$ of individual bunches (solid lines) and wakefield potential $\Phi$ (dotted lines of the same color) at $z=z_m$ for different loads of the reference driver beam: no electrons (``no load''), witness bunches only (``no phs''), witness bunches and optimum phasers (``base''), and witness bunches and phasers with doubled charge (``over''). }
\label{fig15-bunches}
\end{figure}

\begin{figure}[tb]
\includegraphics{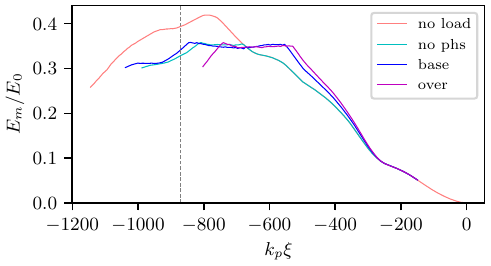}
\caption{ Average electric field amplitude $E_m$ for the same variants as in Fig.\,\ref{fig15-bunches}. }
\label{fig16-overphaser}
\end{figure}

If the wave phase shifts backward due to the witnesses and phasers, the driver bunches at the rear of the train become partially destroyed (Fig.\,\ref{fig15-bunches}).
An excessive phase shift results in a significant loss of driver charge, so there is an optimal phase shift at which the bunches are most efficient (Fig.\,\ref{fig16-overphaser}).
In all cases shown in Figs.\,\ref{fig15-bunches} and~\ref{fig16-overphaser}, the witness bunches, if any, are initially (at $z=z_w$) distributed according to the algorithm described in Sec.~\ref{s-refcase}.
The phasers with doubled charge are twice as long as in the reference case: their length is $2\Delta \xi$.
With no phasers, the wakefield phase shifts slightly under the influence of the witness bunches, and the driver bunches remain almost unchanged and unable to replenish the wakefield (cyan lines in Figs.\,\ref{fig15-bunches} and~\ref{fig16-overphaser}).
The accelerated witness charge (0.44\,nC) in this case is 40\% of the charge in the reference case (blue lines).
If the phaser charge exceeds the optimum value (magenta lines, doubled phaser charge), the witness charge (0.66\,nC) is 60\%  of that in the reference case.
The witness charge correlates with the length of the interval in which the wakefield amplitude remains at the threshold value $E_a$ (Fig.\,\ref{fig16-overphaser}).

\begin{figure}[tb]
\includegraphics{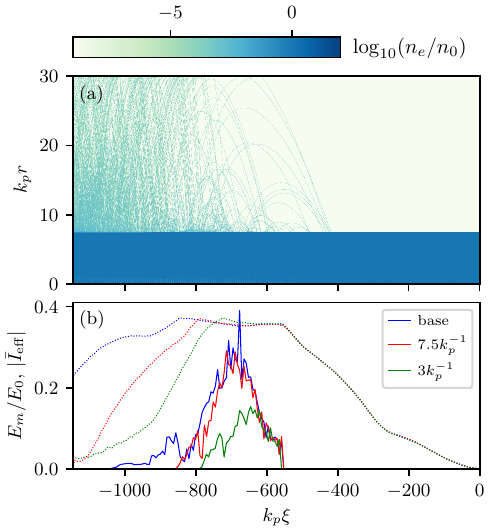}
\caption{ (a) Plasma electron density $n_e (r, \xi)$ at $z=z_m$ for the radially limited plasma of radius $7.5 k_p^{-1}$.
(b) Averaged effective current $\bar I_\text{eff}$ of the witness bunches at $z=z_m$ (solid lines) and the corresponding amplitude of the electric field $E_m$ (dotted lines of the same color) for the reference case (``base'') and radially limited plasmas of radius $7.5 k_p^{-1}$ and $3 k_p^{-1}$. }
\label{fig17-wide}
\end{figure}

\subsection{No vacuum or neutral gas around the plasma}

If the plasma operates near the maximum wakefield amplitudes achievable with multi-bunch excitation, the wave periodically breaks transversely.
This is an event in which the trajectories of plasma electrons, initially at different radii, intersect, causing some electrons to drop out of the collective motion in the wave \cite{PRL78-4205, PoP25-103103, PoP23-103112, PPCF63-055002, PoP29-023104, PoP32-023905}.
The wavebreaking does not result in a rapid decrease in the wave amplitude, but generates jets of fast electrons that diverge radially from the wavebreaking points [Fig.\,\ref{fig17-wide}(a)] \cite{PoP31-034503, PPCF63-055002, PPCF64-045003}.
If the plasma is very wide in the transverse direction, the fast electrons travel freely far away and do not return to the wave region.
When the plasma is bounded by a solid wall, the situation is similar for the wave: the electrons strike the wall and are absorbed.
Secondary electrons, if any, have much lower energies and cannot reach the wave region while the witness bunches pass by.
This second option is implemented in LCODE in such a way that each incident electron produces one low-energy secondary electron, thereby ensuring the integral neutrality of the plasma.

However, if the plasma is surrounded by a region of vacuum or neutral gas, the escape of fast electrons makes the plasma positively charged \cite{PRL112-194801}.
The radial electric field generated around the plasma causes the fast electrons to return to the plasma [Fig.\,\ref{fig17-wide}(a)] \cite{PPCF62-125023, PRAB24-011301}.
The returning electrons predominantly fall into the accelerating phase of the plasma wave, thereby effectively damping it \cite{PPCF63-055002}.
As a consequence, the total accelerated charge is lower.
The charge deficit appears mainly at the rear of the witness bunch train, where the effect of the returning electrons is strongest [Fig.\,\ref{fig17-wide}(b)].
For the plasma of radius $R_p = 7.5k_p^{-1}$ surrounded by a wide vacuum region (limited by the simulation domain boundary at $30 k_p^{-1}$), with all other parameters identical to the reference case, the total witness charge is 0.88\,nC.
This is 20\% lower than in the case with no vacuum region.
For $R_p = 3k_p^{-1}$, the total witness charge is 0.48\,nC, which is less than half that in the reference case.

\begin{figure}[tb]
\includegraphics{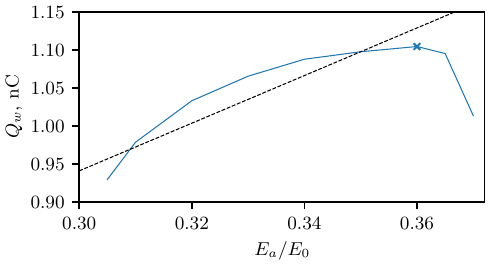}
\caption{ Dependence of the accelerated charge $Q_w$ on the threshold amplitude $E_a$, up to which the wave is allowed to grow.
The slope of the dashed lines shows the proportionality $Q_w \propto E_a$.
The cross marks the reference case.  }
\label{fig18-maxfield}
\end{figure}

\subsection{Witness bunches not at field maxima}

According to Eqs.~\eqref{e2} or~\eqref{e3}, the larger the ratio $E_\text{max}/E_w$, the higher the witness charge.
If the accelerating field $E_w$ is fixed, a higher wakefield amplitude $E_m$ causes each driver bunch to transfer more energy to the plasma wave, proportionally to $E_m$ (Fig.\ref{fig18-maxfield}).
Although the wakefield amplitude is limited by nonlinear effects, there may be an interval between $E_w$ and this limit, and the maximum witness charge corresponds to the upper bound of this interval.
Operating near the lower bound additionally reduces the witness charge (Fig.\ref{fig18-maxfield}), because if a witness bunch is near the field maximum, it is also located near the transition between the focusing and defocusing radial fields, which allows small fluctuations of the wakefield phase to cause bunch defocusing.

In other words, the witnesses should not be injected into the wave as soon as its amplitude slightly exceeds $E_w$.
Instead, the wave should be allowed to grow as much as possible.

\subsection{Additional notes}

In the reference case, the witnesses and phasers are placed in each wakefield period.
This is not necessary for maximizing the charge.
Placing witnesses and phasers with a fivefold charge in every fifth wakefield period does not result in a noticeable decrease in the total witness charge.

\begin{figure}[tb]
\includegraphics{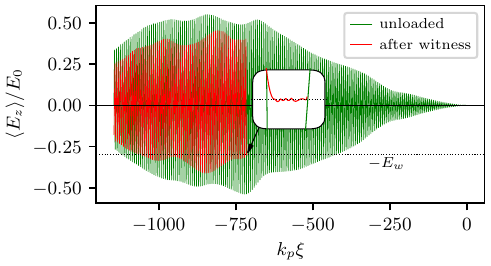}
\caption{ Average electric field $\langle E_z \rangle$ in the variant optimized for accelerating a single large witness, with and without the witness.
The dotted line indicates the required accelerating field $(-E_w)$.  }
\label{fig19-single}
\end{figure}

It is interesting to compare the charge accelerated in multiple witness bunches and in a single large witness bunch using the same initial driver beam with the same leading-edge cutoff at $\xi_c = 2 \sigma_{zb}$.
The plasma density profile for the latter case, optimized to achieve the maximum wakefield amplitude at $z = z_w$, consists of a linear increase from $n_0$ to $1.053 n_0$ over the first 2.2\,m, followed by a constant density of $1.053 n_0$.
The witness is placed at the point of maximum $E_m$, and its longitudinal shape is optimized such that $\langle E_z \rangle = E_w$ within the witness (Fig.\,\ref{fig19-single}).
The resulting witness charge is 0.75\,nC.
The ratio of this charge to the charge accelerated in the reference case roughly equals the ratio of the charge required to drive the wave to the maximum amplitude to the total driver charge in the reference case.
The additional charge that replenishes the wave provides the corresponding increase in the total witness charge.

Let us also compare the charge accelerated at a lower rate $E_w = 0.15 E_0$ with that obtained in Ref.~\cite{PRR7-L012055} for almost the same initial driver beam and the same accelerating field.
If we follow all of the above steps for maximizing the charge, the total witness charge is 2.65\,nC, which is nearly three times the value reported in Ref.~\cite{PRR7-L012055} (0.912\,nC).
If we begin to inject the witness bunches earlier, when the field amplitude reaches only $E_m = 0.18 E_0$, and do not inject phasers, the total witness charge is 1.63\,nC.
Therefore, at low accelerating fields, the factors described in Section~\ref{s5-optimized} become even more significant.

\section{Summary and discussion}
\label{s6-summary}

It is possible to accelerate a high-charge train of short witness bunches in a plasma wave driven by a train of short driver bunches.
In this case, the driver bunches replenish the wave energy after each witness bunch.
The efficient energy exchange, mediated by the plasma, can persist over many wakefield periods until the wave degrades due to ion motion or other mechanisms.
As a result, the witness charge can be quite high --- comparable to that of the driver.

However, implementing this method is complicated by the transverse dynamics of the bunches.
Under the influence of plasma fields, the bunches reshape to reach radial equilibrium with the wave, which in turn modifies the excited wakefield.
This effect is more pronounced for the driver bunches, which typically have a width on the order of the plasma skin depth $k_p^{-1}$.
High-quality witness bunches have a radius well below $k_p^{-1}$, and their interaction with the wakefield is largely insensitive to variations in this radius.
The established equilibrium bunch shape, which is matched to the focusing force exerted by the plasma, is rather exotic.
Reproducing this profile before the beam enters the plasma is hardly feasible.

The straightforward solution is to allow the driver bunches to stabilize before injecting the witnesses.
However, this approach breaks down at high accelerating fields.
If the plasma wave loaded with witness bunches has a high amplitude, the unloaded wave must have an even higher one, at which nonlinear elongation of the wave period leads to beam disruption and wavebreaking.
Thus, the required driver beam cannot propagate through the plasma in the absence of the witness bunches.
It must evolve toward the equilibrium state in the presence of the witness bunches within the wakefield. These experience wakefield evolution, potentially degrading their quality.

We propose to overcome the problem by allowing the driver to evolve into an equilibrium state in which it excites a wakefield of sufficiently low amplitude due to the low efficiency of the formed bunches.
When the witness bunches are injected, they induce a phase shift in the wakefield, forcing the driver bunches to evolve toward a new equilibrium state.
In this state, the driver bunches excite the wave more efficiently.
However, since the wave is already loaded by the witness bunches, its amplitude remains below the limit.
As the initial and final equilibrium states of the driver bunches differ only slightly --- unlike the drastic evolution undergone by the beam upon entering the plasma --- the transition does not significantly perturb the wakefield structure, allowing the witness bunches to preserve their quality.

We have demonstrated this method using a proton beam with parameters matching those of the AWAKE experiment.
The beam evolves into the train of equilibrium bunches as a result of seeded self-modulation.
This realistically formed bunch train is inferior to the idealized case without self-modulation constraints, and the total charge of the witness beam achievable with it falls several times below the theoretical limit.
However, it allows us to identify several factors that can increase the total witness charge in this scheme: optimizing the plasma density profile during the self-modulation stage, injecting additional phaser bunches at the zero-accelerating-field points, suppressing the formation of an electron halo around the plasma, and operating at the highest possible wakefield amplitudes.

This method is also applicable to a pre-bunched driver beam.
In this case, the ratio of the witness charge to the initial driver charge is expected to be even higher.
This should be investigated in future work, preferably based on realistic beam parameters of an existing or proposed facility.

\acknowledgements

Numerical simulations for this work were performed on HPC-cluster ``Akademik V.\,M.\,Matrosov'' \cite{Matrosov}.



\begin{thebibliography}{30}
\bibitem{CAS-2019}
    Proceedings of the 2019 CERN–Accelerator–School course on High Gradient Wakefiled Accelerators, Sesimbra (Portugal), CERN (2020), \href{http://dx.doi.org/10.17181/CAS2019Sesimbra}{http://dx.doi.org/10.17181/CAS2019Sesimbra}.
\bibitem{RMP81-1229}
   E. Esarey, C.B. Schroeder, and W.P. Leemans,
   Physics of laser-driven plasma-based electron accelerators.
   \href{http://dx.doi.org/10.1103/RevModPhys.81.1229}{Rev. Mod. Phys.} \textbf{81}, 1229 (2009).
\bibitem{IEEE-PS24-252}
    E. Esarey, P. Sprangle, J. Krall, and A. Ting,
    Overview of plasma-based accelerator concepts.
    \href{http://dx.doi.org/10.1109/27.509991}{IEEE Trans. Plasma Sci.} \textbf{24}, 252 (1996).
\bibitem{MRE9-014001}
    C. Aniculaesei, T. Ha, S. Yoffe, L. Labun, S. Milton, E. McCary, M.M. Spinks, H.J. Quevedo, O.Z. Labun, R. Sain, et al., 
    The acceleration of a high-charge electron bunch to 10 GeV in a 10-cm nanoparticle-assisted wakefield accelerator.
    \href{https://doi.org/10.1063/5.0161687}{Matter Radiat. Extremes} \textbf{9}, 014001 (2024).
\bibitem{PRL133-255001}
    A. Picksley, J. Stackhouse, C. Benedetti, K. Nakamura, H.E. Tsai, R. Li, B. Miao, J.E. Shrock, E. Rockafellow, H.M. Milchberg, et al., 
    Matched Guiding and Controlled Injection in Dark-Current-Free, 10-GeV-Class, Channel-Guided Laser-Plasma Accelerators.
    \href{https://doi.org/10.1103/PhysRevLett.133.255001}{Phys. Rev. Lett.} \textbf{133}, 255001 (2024).
\bibitem{PPCF58-034017}
    M. Litos, E. Adli, J.M. Allen, W. An, C.I. Clarke, S. Corde, C.E. Clayton, J. Frederico, S.J. Gessner, S.Z. Green, et al., 
    9 GeV energy gain in a beam-driven plasma wakefield accelerator.
    \href{http://dx.doi.org/10.1088/0741-3335/58/3/034017}{Plasma Phys. Control. Fusion} \textbf{58}, 034017 (2016).
\bibitem{PRL126-214801}
    L.T. Ke, K. Feng, W.T. Wang, Z.Y. Qin, C.H. Yu, Y. Wu, Y. Chen, R. Qi, Z.J. Zhang, Y. Xu, X.J. Yang, Y.X. Leng, J.S. Liu, R.X. Li, and Z.Z. Xu,
    Near-GeV Electron Beams at a Few Per-Mille Level from a Laser Wakefield Accelerator via Density-Tailored Plasma.
    \href{https://doi.org/10.1103/PhysRevLett.126.214801}{Phys. Rev. Lett.} \textbf{126}, 214801 (2021).
\bibitem{Nat.640-907}
    P. Winkler, M. Trunk, L. H\"ubner, A. Martinez de la Ossa, S. Jalas, M. Kirchen, I. Agapov, S.A. Antipov, R. Brinkmann, T. Eichner, et al., 
    Active energy compression of a laser-plasma electron beam.
    \href{https://doi.org/10.1038/s41586-025-08772-y}{Nature} \textbf{640}, 907 (2025).
\bibitem{NJP23-031101}
    F. Albert, M.E. Couprie, A. Debus, M.C. Downer, J. Faure, A. Flacco, L.A. Gizzi, T. Grismayer, A. Huebl, C. Joshi, et al.,
    2020 roadmap on plasma accelerators.
    \href{https://doi.org/10.1088/1367-2630/abcc62}{New J. Phys.} \textbf{23}, 031101 (2021).
\bibitem{PoP27-070602}
    C. Joshi, S. Corde, and W.B. Mori,
    Perspectives on the generation of electron beams from plasma-based accelerators and their near and long term applications.
    \href{https://doi.org/10.1063/5.0004039}{Phys. Plasmas} \textbf{27}, 070602 (2020).
\bibitem{PRR7-023118}
    T.N. Dalichaouch, X.L. Xu, F. Li, F.S. Tsung, and W.B. Mori,
    Fully plasma-based electron injector for a linear collider or XFEL.
    \href{https://doi.org/10.1103/PhysRevResearch.7.023118}{Phys. Rev. Res.} \textbf{7}, 023118 (2025).
\bibitem{PPCF63-125027}
    K.V. Lotov and P.V. Tuev,
    Plasma wakefield acceleration beyond the dephasing limit with 400 GeV proton driver.
    \href{https://doi.org/10.1088/1361-6587/ac349a}{Plasma Phys. Control. Fusion} \textbf{63}, 125027 (2021).
\bibitem{BLPI50-S715}
    M.S. Dorozhkina, K.V. Baluev, D.D. Kutergin, I.K. Lotov, V.A. Minakov, R.I. Spitsyn, P.V. Tuev, and K.V. Lotov,
    Laser Wakefield Acceleration in a Plasma Channel.
    \href{https://doi.org/10.3103/S1068335623180057}{Bulletin of the Lebedev Physics Institute} \textbf{50}, S715 (2023).
\bibitem{PRL43-267}
    T. Tajima and J.M. Dawson,
    Laser Electron Accelerator.
    \href{https://doi.org/10.1103/PhysRevLett.43.267}{Phys. Rev. Lett.} \textbf{43}, 267 (1979).
\bibitem{PRL54-693}
    P. Chen, J.M. Dawson, R.W. Huff, and T. Katsouleas,
    Acceleration of Electrons by the~Interaction of a Bunched Electron Beam with a Plasma.
    \href{ https://doi.org/10.1103/PhysRevLett.54.693}{Phys. Rev. Lett.} \textbf{54}, 693 (1985); \href{https://doi.org/10.1103/PhysRevLett.55.1537}{Phys. Rev. Lett.} \textbf{55}, 1537 (1985).
\bibitem{PRL104-255003}
	N. Kumar, A. Pukhov, and K. Lotov,
	Self-modulation instability of a long proton bunch in plasmas.
	\href{http://dx.doi.org/10.1103/PhysRevLett.104.255003}{Phys. Rev. Lett.} \textbf{104}, 255003 (2010).
\bibitem{JPB47-234003}
    S.M. Hooker, R. Bartolini, S.P.D. Mangles, A. T\"unnermann, L. Corner, J. Limpert, A. Seryi, and R. Walczak,
    Multi-pulse laser wakefield acceleration: a new route to efficient, high-repetition-rate plasma accelerators and high flux radiation sources.
    \href{http://dx.doi.org/10.1088/0953-4075/47/23/234003}{J. Phys. B: At. Mol. Opt. Phys.} \textbf{47}, 234003 (2014).
\bibitem{IPAC10-4395}
	G. Xia, A. Caldwell,
	Producing Short Proton Bunch For Driving Plasma Wakefield Acceleration.
    \href{https://proceedings.jacow.org/IPAC10/papers/thpd051.pdf}{Proc. IPAC2010} (Kyoto, Japan), pp.~4395--4397.
\bibitem{PPCF53-014003}
	A. Caldwell, K. Lotov, A. Pukhov and G. Xia,
	Plasma wakefield excitation with a 24 GeV proton beam.
    \href{http://dx.doi.org/10.1088/0741-3335/53/1/014003}{Plasma Phys. Controlled Fusion} \textbf{53}, 014003 (2011).
\bibitem{PRST-AB7-111301}
    R. Maeda, T. Katsouleas, P. Muggli, C. Joshi, W.B. Mori, and W. Quillinan,
    Possibility of a multibunch plasma afterburner for linear colliders.
    \href{https://doi.org/10.1103/PhysRevSTAB.7.111301}{Phys. Rev. ST Accel. Beams} \textbf{7}, 111301 (2004).
\bibitem{PRR7-L012055}
    J.P. Farmer, G. Zevi Della Porta,
    Wakefield regeneration in a plasma accelerator.
    \href{https://doi.org/10.1103/PhysRevResearch.7.L012055}{Phys. Rev. Research} \textbf{7}, L012055 (2025).
\bibitem{PoP20-083119}
    K.V. Lotov,
    Excitation of two-dimensional plasma wakefields by trains of equidistant particle bunches.
    \href{http://dx.doi.org/10.1063/1.4819720}{Phys. Plasmas} \textbf{20}, 083119 (2013).
\bibitem{PAcc17-171}
    R.D. Ruth, A.W. Chao, P.L. Morton, and P.B. Wilson,
    A Plasma Wake-Field Accelerator.
    \href{https://inspirehep.net/literature/206328}{Part. Accel.} \textbf{17}, 171 (1985).
\bibitem{AIP396-75}
    B.N. Breizman, P.Z. Chebotaev, A.M. Kudryavtsev, K.V. Lotov, and A.N. Skrinsky,
    Self-Focused Particle Beam Drivers for Plasma Wakefield Accelerators.
    \href{https://doi.org/10.1063/1.52975}{AIP Conf. Proc.} \textbf{396}, 75 (1997).
\bibitem{RMP94-025006}
    G. Ha , K.-J. Kim, J.G. Power, Y. Sun, and P. Piot,
    Bunch shaping in electron linear accelerators.
    \href{https://doi.org/10.1103/RevModPhys.94.025006}{Rev. Mod. Phys.} \textbf{94}, 025006 (2022).
\bibitem{JINST17-P05036}
    F. Lemery, G. Andonian, S. Doebert, G. Ha, X. Lu, J. Power and E. Wisniewski,
    Drive beam sources and longitudinal shaping techniques for beam driven accelerators.
    \href{https://doi.org/10.1088/1748-0221/17/05/P05036}{JINST} \textbf{17}, P05036 (2022).
\bibitem{RuPAC16-303}
    I. Sheinman, A. Petrenko,
    High-energy micro-buncher based on the mm-wavelength dielectric structure.
    \href{https://proceedings.jacow.org/rupac2016/papers/tupsa041.pdf}{Proc. RuPAC2016} (St. Petersburg, Russia), pp.~303--306.
\bibitem{PFB5-2714}
    B.L. Militsyn, A.A. Bechtenev, B.N. Breizman, P.Z. Chebotaev, I.A. Koop, A.M. Kudryavtsev, V.M. Panasyuk, Yu.M. Shatunov, and A.N. Skrinsky,
    Experimental plasma wake-field acceleration project.
    \href{https://doi.org/10.1063/1.860710}{Phys.~Fluids~B} \textbf{5}, 2714 (1993).
\bibitem{PPR31-292}
    A.V. Burdakov, A.M. Kudryavtsev, P.V. Logatchov, K.V. Lotov, A.V. Petrenko, and A.N. Skrinsky,
    Design of an Experiment on Wakefield Acceleration on the VEPP-5 Injection Complex.
    \href{https://doi.org/10.1134/1.1904145}{Plasma Physics Reports} \textbf{31}, 292 (2005).
\bibitem{AIP1299-467}
	A.V. Petrenko, K.V. Lotov, P.V. Logatchov and A.V. Burdakov,
	The Facility for 500 MeV Plasma Wake-Field Acceleration Experiments at Budker INP.
	\href{https://doi.org/10.1063/1.3520366}{AIP Conf. Proc.} \textbf{1299}, 467 (2010).
\bibitem{PRST-AB13-052803}
    P. Muggli, B. Allen, V.E. Yakimenko, J. Park, M. Babzien, K.P. Kusche, and W.D. Kimura,
    Simple method for generating adjustable trains of picosecond electron bunches.
    \href{http://dx.doi.org/10.1103/PhysRevSTAB.13.052803}{Phys. Rev. ST Accel. Beams} \textbf{13}, 052803 (2010).
\bibitem{NIMA-916-195}
    J. Hyun, P. Piot, T. Sen, J.C. Thangaraj,
    Micro-bunching for generating tunable narrow-band THz radiation at the FAST photoinjector.
    \href{https://doi.org/10.1016/j.nima.2018.10.219}{Nucl. Instrum. Methods Phys. Res. A} \textbf{916}, 195 (2019).
\bibitem{PRAB23-072803}
    G. Ha, J.G. Power, J. Shao, M. Conde, and C. Jing,
    Coherent synchrotron radiation free longitudinal bunch shaping using transverse deflecting cavities.
    \href{https://doi.org/10.1103/PhysRevAccelBeams.23.072803}{Phys. Rev. ST Accel. Beams} \textbf{23}, 072803 (2020).
\bibitem{PoP22-103110}
	K.V. Lotov,
	Physics of beam self-modulation in plasma wakefield accelerators.
	\href{http://dx.doi.org/10.1063/1.4933129}{Phys. Plasmas} \textbf{22}, 103110 (2015).
\bibitem{PRST-AB6-061301}
    K.V. Lotov,
    Fine wakefield structure in the blowout regime of plasma wakefield accelerators.
    \href{https://doi.org/10.1103/PhysRevSTAB.6.061301}{Phys. Rev. ST Accel. Beams} \textbf{6}, 061301 (2003).
\bibitem{NIMA-829-350}
	A.P. Sosedkin, K.V. Lotov,
	LCODE: A parallel quasistatic code for computationally heavy problems of plasma wakefield acceleration.
    \href{http://dx.doi.org/10.1016/j.nima.2015.12.032}{Nucl. Instrum. Methods Phys. Res. A} \textbf{829}, 350 (2016).
\bibitem{PoP32-033102}
    N.V. Okhotnikov and K.V. Lotov,
    Acceleration rate enhancement by negative plasma density gradient in multi-bunch driven plasma wakefield accelerator.
    \href{https://doi.org/10.1063/5.0249127}{Phys. Plasmas} \textbf{32}, 033102 (2025).
\bibitem{PPCF64-075003}
    M.A. Baistrukov and K.V. Lotov,
    Evolution of equilibrium particle beams in plasma under external wakefields.
    \href{https://doi.org/10.1088/1361-6587/ac6ffe}{Plasma Phys. Controlled Fusion} \textbf{64}, 075003 (2022).
\bibitem{PoP5-785}
    K.V. Lotov,
    Simulation of ultrarelativistic beam dynamics in plasma wake-field accelerator.
    \href{https://doi.org/10.1063/1.872765}{Phys. Plasmas} \textbf{5}, 785 (1998).
\bibitem{NIMA-410-461}
    K.V. Lotov,
    Simulation of ultrarelativistic beam dynamics in the plasma wake-field accelerator.
    \href{https://doi.org/10.1016/S0168-9002(98)00178-8}{Nucl. Instrum. Methods Phys. Res. A} \textbf{410}, 461 (1998).
\bibitem{PoP24-023119}
    K.V. Lotov,
    Radial equilibrium of relativistic particle bunches in plasma wakefield accelerators.
    \href{http://dx.doi.org/10.1063/1.4977058}{Phys. Plasmas} \textbf{24}, 023119 (2017).
\bibitem{features}
	V.M. Yarygova and K.V. Lotov,
	Features of long particle beam self-modulated in plasma.
	\href{https://arxiv.org/abs/2606.25663}{arXiv:2606.25663} [physics.acc-ph].
\bibitem{PPCF60-024002}
    K.V. Lotov,
    Stable bunch trains for plasma wakefield acceleration.
    \href{https://doi.org/10.1088/1361-6587/aa9f97}{Plasma Phys. Controlled Fusion} \textbf{60}, 024002 (2018).
\bibitem{EPAC98-806}
    K.V. Lotov,
    Instability of long driving beams in plasma wakefield accelerators.
    (\href{https://jacow.org/e98/papers/MOP12E.pdf}{Proc. EPAC1998} (Stockholm, 1998), pp.~806-–808.

\bibitem{PRL122-054802}
    E. Adli, A. Ahuja, O. Apsimon, R. Apsimon, A.-M. Bachmann, D. Barrientos, M. M. Barros, J. Batkiewicz, F. Batsch, J. Bauche, et al. (The AWAKE Collaboration),
    Experimental Observation of Proton Bunch Modulation in a Plasma at Varying Plasma Densities.
    \href{https://doi.org/10.1103/PhysRevLett.122.054802}{Phys. Rev. Lett.} \textbf{122}, 054802 (2019).
\bibitem{PRL122-054801}
    M. Turner, E. Adli, A. Ahuja, O. Apsimon, R. Apsimon, A.-M. Bachmann, M. Barros Marin, D. Barrientos, F. Batsch, J. Batkiewicz, et al. (The AWAKE Collaboration),
    Experimental Observation of Plasma Wakefield Growth Driven by the Seeded Self-Modulation of a Proton Bunch.
    \href{https://doi.org/10.1103/PhysRevLett.122.054801}{Phys. Rev. Lett.} \textbf{122}, 054801 (2019).
\bibitem{PRL126-164802}
    F. Batsch, P. Muggli, R. Agnello, C.C. Ahdida, M.C. Amoedo Goncalves, Y. Andrebe, O. Apsimon, R. Apsimon, A.-M. Bachmann, M.A. Baistrukov, et al. (The AWAKE Collaboration),
    Transition between Instability and Seeded Self-Modulation of a Relativistic Particle Bunch in Plasma.
    \href{https://doi.org/10.1103/PhysRevLett.126.164802}{Phys. Rev. Lett.} \textbf{126}, 164802 (2021).
\bibitem{PRAB24-101301}
    P.I. Morales Guzman, P. Muggli, R. Agnello, C.C. Ahdida, M. Aladi, M.C. Amoedo Goncalves, Y. Andrebe, O. Apsimon, R. Apsimon, A.-M. Bachmann, et al. (The AWAKE Collaboration),
    Simulation and experimental study of proton bunch self-modulation in plasma with linear density gradients.
    \href{https://doi.org/10.1103/PhysRevAccelBeams.24.101301}{Phys. Rev. Accel. Beams} \textbf{24}, 101301 (2021).
\bibitem{PRL125-264801}
    F. Braunmuller, T. Nechaeva, E. Adli, R. Agnello, M. Aladi, Y. Andrebe, O. Apsimon, R. Apsimon, A.-M. Bachmann, M.A. Baistrukov, et al. (The AWAKE Collaboration),
    Proton Bunch Self-Modulation in Plasma with Density Gradient.
    \href{https://doi.org/10.1103/PhysRevLett.125.264801}{Phys. Rev. Lett.} \textbf{125}, 264801 (2020).
\bibitem{PPCF56-084014}
    J. Vieira, L.D. Amorim, Y. Fang, W.B. Mori, P. Muggli, and L.O. Silva,
    Self-modulation instability of ultra-relativistic particle bunches with finite rise times,
    \href{http://dx.doi.org/10.1088/0741-3335/56/8/084014}{Plasma Phys. Controlled Fusion} \textbf{56}, 084014 (2014).
\bibitem{PRL112-045001}
    Y. Fang, V. E. Yakimenko, M. Babzien, M. Fedurin, K. P. Kusche, R. Malone, J. Vieira, W. B. Mori, and P. Muggli,
    Seeding of Self-Modulation Instability of a Long Electron Bunch in a Plasma.
    \href{http://dx.doi.org/10.1103/PhysRevLett.112.045001}{Phys. Rev. Lett.} \textbf{112}, 045001 (2014).
\bibitem{PRL120-144802}
    M. Gross, J. Engel, J. Good, H. Huck, I. Isaev, G. Koss, M. Krasilnikov, O. Lishilin, G. Loisch, Y. Renier, et al., 
    Observation of the Self-Modulation Instability via Time-Resolved Measurements.
    \href{https://doi.org/10.1103/PhysRevLett.120.144802}{Phys. Rev. Lett.} \textbf{120}, 144802 (2018).
\bibitem{PPCF61-045012}
    G. Loisch, G. Asova, P. Boonpornprasert, Y. Chen, J. Good, M. Gross, H. Huck, D. Kalantaryan, M. Krasilnikov, O. Lishilin, D. Melkumyan, A. Oppelt, H. Qian, Y. Renier, F. Stephan and Q. Zhao,
    Plasma density measurement by means of self-modulation of long electron bunches.
    \href{https://doi.org/10.1088/1361-6587/ab04b9}{Plasma Phys. Controlled Fusion} \textbf{61}, 045012 (2019).
\bibitem{PPCF62-115025}
    K.V. Lotov and V.A. Minakov,
    Proton beam self-modulation seeded by electron bunch in plasma with density ramp.
    \href{https://doi.org/10.1088/1361-6587/abba42}{Plasma Phys. Controlled Fusion} \textbf{62}, 115025 (2020).
\bibitem{PRL129-024802}
    L. Verra, G. Zevi Della Porta, J. Pucek, T. Nechaeva, S. Wyler, M. Bergamaschi, E. Senes, E. Guran, J.T. Moody, M.A. Kedves, et al. (The AWAKE Collaboration),
    Controlled Growth of the Self-Modulation of a Relativistic Proton Bunch in Plasma.
    \href{https://doi.org/10.1103/PhysRevLett.129.024802}{Phys. Rev. Lett.} \textbf{129}, 024802 (2022).
\bibitem{PoP20-103111}
    C. Siemon, V. Khudik, S.A. Yi, A. Pukhov, and G. Shvets,
    Laser-seeded modulation instability in a proton driver plasma wakefield accelerator.
    \href{http://dx.doi.org/10.1063/1.4825245}{Phys. Plasmas} \textbf{20}, 103111 (2013).
\bibitem{PoP18-024501}
	K.V. Lotov,
	Controlled self-modulation of high energy beams in a plasma.
	\href{http://dx.doi.org/10.1063/1.3558697}{Phys. Plasmas} \textbf{18}, 024501 (2011).
\bibitem{PoP18-103101}
	A. Caldwell, K.V. Lotov,
	Plasma wakefield acceleration with a modulated proton bunch.
	\href{http://dx.doi.org/10.1063/1.3641973}{Phys. Plasmas} \textbf{18}, 103101 (2011).
\bibitem{Symmetry14-1680}
    E. Gschwendtner, K. Lotov, P. Muggli, M. Wing, R. Agnello, C.C. Ahdida, M.C. Amoedo Goncalves, Y. Andrebe, O. Apsimon, R. Apsimon, et al. (The AWAKE Collaboration),
    The AWAKE Run 2 programme and beyond.
    \href{https://doi.org/10.3390/sym14081680}{Symmetry} \textbf{14}, 1680 (2022).
\bibitem{NIMA-829-63}
	A. Petrenko, K. Lotov, A. Sosedkin,
    Numerical Studies of Electron Acceleration Behind Self-Modulating Proton Beam in Plasma with a Density Gradient.
    \href{http://dx.doi.org/10.1016/j.nima.2016.01.063}{Nucl. Instrum. Methods Phys. Res. A} \textbf{829}, 63 (2016).
\bibitem{Nat.561-363}
    E. Adli, A. Ahuja, O. Apsimon, R. Apsimon, A.-M. Bachmann, D. Barrientos, F. Batsch, J. Bauche, V.K. Berglyd Olsen, M. Bernardini et al. (The AWAKE Collaboration),
    Acceleration of electrons in the plasma wakefield of a proton bunch.
    \href{https://doi.org/10.1038/s41586-018-0485-4}{Nature} \textbf{561}, 363 (2018).
\bibitem{PPCF62-125023}
    A.A. Gorn, M. Turner, E. Adli, R. Agnello, M. Aladi, Y. Andrebe, O. Apsimon, R. Apsimon, A.-M. Bachmann, M.A. Baistrukov, et al. (The AWAKE Collaboration),
    Proton beam defocusing in AWAKE: comparison of simulations and measurements.
    \href{https://doi.org/10.1088/1361-6587/abc298}{Plasma Phys. Controlled Fusion} \textbf{62}, 125023 (2020).
\bibitem{PRL107-145003}
    A. Pukhov, N. Kumar, T. Tuckmantel, A. Upadhyay, K. Lotov, P. Muggli, V. Khudik, C. Siemon, and G. Shvets,
	Phase Velocity and Particle Injection in a Self-Modulated Proton-Driven Plasma Wakefield Accelerator.
	\href{http://dx.doi.org/10.1103/PhysRevLett.107.145003}{Phys. Rev. Lett.} \textbf{107}, 145003 (2011).
\bibitem{PoP19-010703}
    C.B. Schroeder, C. Benedetti, E. Esarey, F.J. Gruener, and W.P. Leemans,
    Particle beam self-modulation instability in tapered and inhomogeneous plasma.
    \href{http://dx.doi.org/10.1063/1.3677358}{Phys. Plasmas} \textbf{19}, 010703 (2012).
\bibitem{JPCS1067-042009}
    Y. Li, G. Xia, K.V. Lotov, A.P. Sosedkin, Y. Zhao, S. J.Gessner,
    Amplitude enhancement of the self-modulated plasma wakefields.
    \href{https://doi.org/10.1088/1742-6596/1067/4/042009}{Journal of Physics: Conf. Series} \textbf{1067}, 042009 (2018).
\bibitem{PRL134-155001}
    M. Turner, E. Walter, C. Amoedo, N. Torrado, N. Lopes, A. Sublet, M. Bergamaschi, J. Pucek, J. Mezger, N. van Gils, et al. (The AWAKE Collaboration),
    Experimental Observation of the Motion of Ions in a Resonantly Driven Plasma Wakefield Accelerator.
    \href{https://doi.org/10.1103/PhysRevLett.134.155001}{Phys. Rev. Lett.} \textbf{134}, 155001 (2025).
\bibitem{2602.16624}
    A. Clairembaud, M. Turner, M. Bergamaschi, L. Ranc, F. Pannell, J. Mezger, H. Jaworska, N. van Gils, J. Farmer, P. Muggli, et al. (The AWAKE Collaboration),
	Measurement of the Saturation Length of the Self-Modulation Instability.
	\href{https://arxiv.org/abs/2602.16624}{arXiv:2602.16624} [physics.plasm-ph].
\bibitem{PRAB29-083601}
    N.Z. van Gils, E. Belli, M. Bergamaschi, A. Clairembaud, A. Gerbershagen, E. Gschwendtner, H. Jaworska, J. Mezger, M. Moreira, P. Muggli, et al. (The AWAKE Collaboration),
    Seeding of self-modulation using truncated seed bunches as a path to high gradient acceleration.
    \href{https://doi.org/10.1103/c7jm-hf4q}{Phys. Rev. Accel. Beams} \textbf{29}, 083601 (2026).
\bibitem{NIMA-829-3}
    A. Caldwell, E. Adli, L. Amorim, R. Apsimon, T. Argyropoulos, R. Assmann, A.-M. Bachmann, F. Batsch, J. Bauche, V.K. Berglyd Olsen, et al. (The AWAKE Collaboration),
	Path to AWAKE: Evolution of the concept.
    \href{http://dx.doi.org/10.1016/j.nima.2015.12.050}{Nucl. Instrum. Methods Phys. Res. A} \textbf{829}, 3 (2016).
\bibitem{NIMA-829-76}
    E. Gschwendtner, E. Adli, L. Amorim, R. Apsimon, R. Assmann, A.-M. Bachmann, F. Batsch, J. Bauche, V.K. Berglyd Olsen, M. Bernardini, et al.,
	AWAKE, The Advanced Proton Driven Plasma Wakefield Acceleration Experiment at CERN.
    \href{http://dx.doi.org/10.1016/j.nima.2016.02.026}{Nucl. Instrum. Methods Phys. Res. A} \textbf{829}, 76 (2016).
\bibitem{SPSC-SR-337}
    AWAKE Collaboration,
    AWAKE status report.
    \href{https://cds.cern.ch/record/2878573/files/SPSC-SR-337.pdf}{CERN-SPSC-2023-032 / SPSC-SR-337} (2023).
\bibitem{JPD51-025203}
    G. Plyushchev, R. Kersevan, A. Petrenko, and P. Muggli,
    A rubidium vapor source for a plasma source for AWAKE.
    \href{https://doi.org/10.1088/1361-6463/aa9dd7}{J. Phys. D: Appl. Phys.} \textbf{51}, 025203 (2018).
\bibitem{PRAB21-011301}
    V.K. Berglyd Olsen, E. Adli, and P. Muggli,
    Emittance preservation of an electron beam in a loaded quasilinear plasma wakefield.
    \href{https://doi.org/10.1103/PhysRevAccelBeams.21.011301}{Phys. Rev. Accel. Beams} \textbf{21}, 011301 (2018).
\bibitem{PAcc22-81}
    T. Katsouleas, S. Wilks, P. Chen, J.M. Dawson, and J.J. Su,
    Beam loading in plasma accelerators.
    \href{https://inspirehep.net/literature/253298}{Part. Accel.} \textbf{22}, 81 (1987).
\bibitem{PoP12-053105}
    K.V. Lotov,
    Efficient operating mode of the plasma wakefield accelerator.
    \href{https://doi.org/10.1063/1.1889444}{Phys. Plasmas} \textbf{12}, 053105 (2005).
\bibitem{PoP16-056705}
    M. Tzoufras, W. Lu, F.S. Tsung, C. Huang, W.B. Mori, T. Katsouleas, J. Vieira, R.A. Fonseca, and L.O. Silva,
    Beam loading by electrons in nonlinear plasma wakes.
    \href{http://dx.doi.org/10.1063/1.3118628}{Phys. Plasmas} \textbf{16}, 056705 (2009).
\bibitem{PoP30-053108}
    Q. Su, J. Larson, T.N. Dalichaouch, F. Li, W. An, L. Hildebrand, Y. Zhao, V. Decyk, P. Alves, S M. Wild, and W.B. Mori,
    Optimization of transformer ratio and beam loading in a plasma wakefield accelerator with a structure-exploiting algorithm.
    \href{https://doi.org/10.1063/5.0142940}{Phys. Plasmas} \textbf{30}, 053108 (2023).
\bibitem{PPNL21-316}
    D.D. Kutergin, I.K. Lotov, V.A. Minakov, R.I. Spitsyn, P.V. Tuev, and K.V. Lotov,
    Plasma Wakefield Acceleration Driven by XCELS Laser Pulse.
    \href{https://doi.org/10.1134/S1547477124700183}{Physics of Particles and Nuclei Letters} \textbf{21}, 316 (2024).
\bibitem{PoP21-123116}
    K.V. Lotov, A.P. Sosedkin, A.V. Petrenko, L.D. Amorim, J. Vieira, R.A. Fonseca, L.O. Silva, E. Gschwendtner, and P. Muggli,
	Electron trapping and acceleration by the plasma wakefield of a self-modulating proton beam.
	\href{http://dx.doi.org/10.1063/1.4904365}{Phys. Plasmas} \textbf{21}, 123116 (2014).
\bibitem{PRAB24-011301}
    J. Chappell, E. Adli, R. Agnello, M. Aladi, Y. Andrebe, O. Apsimon,  R. Apsimon, A.-M. Bachmann,  M.A. Baistrukov,  F. Batsch, et al. (The AWAKE Collaboration),
    Experimental study of extended timescale dynamics of a plasma wakefield driven by a self-modulated proton bunch.
    \href{https://doi.org/10.1103/PhysRevAccelBeams.24.011301}{Phys. Rev. Accel. Beams} \textbf{24}, 011301 (2021).
\bibitem{PRAB22-031301}
    M. Moreira, J. Vieira, and P. Muggli,
    Influence of proton bunch parameters on a proton-driven plasma wakefield acceleration experiment.
    \href{https://doi.org/10.1103/PhysRevAccelBeams.22.031301}{Phys. Rev. Accel. Beams} \textbf{22}, 031301 (2019).
\bibitem{PRL78-4205}
    S.V. Bulanov, F. Pegoraro, A.M. Pukhov, and A.S. Sakharov,
    Transverse-wake wave breaking.
    \href{https://doi.org/10.1103/PhysRevLett.78.4205}{Phys. Rev. Lett.} \textbf{78}, 4205 (1997).
\bibitem{PoP25-103103}
    R.I. Spitsyn, I.V. Timofeev, A.P. Sosedkin, and K.V. Lotov,
    Characterization of wavebreaking time and dissipation of weakly nonlinear wakefields due to ion motion.
    \href{https://doi.org/10.1063/1.5048549}{Phys. Plasmas} \textbf{25}, 103103 (2018).
\bibitem{PoP23-103112}
    J. Luo, M. Chen, G.-B. Zhang, T. Yuan, J.-Y. Yu, Z.-C. Shen, L.-L. Yu, S.-M. Weng, C.B. Schroeder, and E. Esarey,
    Dynamics of boundary layer electrons around a laser wakefield bubble.
    \href{http://dx.doi.org/10.1063/1.4966047}{Phys. Plasmas} \textbf{23}, 103112 (2016).
\bibitem{PPCF63-055002}
    R.I. Spitsyn and K.V. Lotov,
    Wakefield decay in a radially bounded plasma due to formation of electron halo.
    \href{https://doi.org/10.1088/1361-6587/abe055}{Plasma Phys. Controlled Fusion} \textbf{63}, 055002 (2021).
\bibitem{PoP29-023104}
    A.A. Gorn and K.V. Lotov,
    Generation of plasma electron halo by a charged particle beam in a low density plasma.
    \href{https://doi.org/10.1063/5.0080675}{Phys. Plasmas} \textbf{29}, 023104 (2022).
\bibitem{PoP32-023905}
    I.Yu. Kargapolov, N.V. Okhotnikov, I.A. Shalimova, A.P. Sosedkin, and K.V. Lotov,
    Declustering of macroparticles in long-term simulations of plasma wakefield acceleration.
    \href{https://doi.org/10.1063/5.0251688}{Phys. Plasmas} \textbf{32}, 023905 (2025).
\bibitem{PoP31-034503}
    V.A. Minakov and K.V. Lotov,
    Transformer ratio growth due to ion motion in plasma wakefield accelerators.
    \href{https://doi.org/10.1063/5.0197176}{Phys. Plasmas} \textbf{31}, 034503 (2024).
\bibitem{PPCF64-045003}
    V.K. Khudiakov, K.V. Lotov and M.C. Downer,
    Ion dynamics driven by a strongly nonlinear plasma wake.
    \href{https://doi.org/10.1088/1361-6587/ac4523}{Plasma Phys. Controlled Fusion} \textbf{64}, 045003 (2022).
\bibitem{PRL112-194801}
	K.V. Lotov, A.P. Sosedkin, A.V. Petrenko,
	Long-Term Evolution of Broken Wakefields in Finite-Radius Plasmas.
	\href{http://dx.doi.org/10.1103/PhysRevLett.112.194801}{Phys. Rev. Lett.} \textbf{112}, 194801 (2014).
\bibitem{Matrosov}
    Irkutsk Supercomputer Center of SB RAS, \href{http://hpc.icc.ru}{http://hpc.icc.ru}.

\end{thebibliography}
\end{document}